\documentclass{article}
\pdfoutput=1
\usepackage[final, nonatbib]{nips_2017}
\usepackage[utf8]{inputenc}
\usepackage{amsmath,amssymb,graphicx,romannum}
\usepackage{url}
\usepackage{color}
\usepackage{mathtools}
\usepackage{cite}
\usepackage{booktabs}
\usepackage{float, soul}
\usepackage{lipsum,adjustbox}
\usepackage{verbatim}
\usepackage{tikz}
\usepackage{pgfplots}
\usepackage{subcaption}
\usepackage{multirow, balance}
\usetikzlibrary{pgfplots.groupplots}
\usetikzlibrary{chains,scopes,positioning,arrows,fit}
\usetikzlibrary{arrows, positioning, patterns, calc, decorations.pathmorphing}
\pgfplotsset{compat=1.14}

\newcommand{\cardinality}{\ensuremath{k}}

\title{Classification vs. Regression in Supervised Learning\\for Single Channel Speaker Count Estimation}

\author{
  Fabian-Robert Stöter \\
  International Audio Laboratories Erlangen\thanks{
      International Audio Laboratories Erlangen is a joint institution of the Friedrich-Alexander-Universität Erlangen-N\"urnberg (FAU) and Fraunhofer Institute for Integrated Circuits (IIS).
  }\\
  \texttt{fabian-robert.stoeter@audiolabs-erlangen.de} \\
  \And
  Soumitro Chakrabarty \\
  International Audio Laboratories Erlangen \\
  \texttt{soumitro.chakrabarty@audiolabs-erlangen.de} \\
  \And
  Bernd Edler \\
  International Audio Laboratories Erlangen \\
  \texttt{bernd.edler@audiolabs-erlangen.de} \\
  \And
  Emanuël A. P. Habets \\
  International Audio Laboratories Erlangen \\
  \texttt{emanuel.habets@audiolabs-erlangen.de}
}

\begin{document}
\maketitle
\begin{abstract}
The task of estimating the maximum number of concurrent speakers from single channel mixtures is important for various audio-based applications, such as blind source separation, speaker diarisation, audio surveillance or auditory scene classification. Building upon powerful machine learning methodology, we develop a Deep Neural Network (DNN) that estimates a speaker count. While DNNs efficiently map input representations to output targets, it remains unclear how to best handle the network output to infer integer source count estimates, as a discrete count estimate can either be tackled as a regression or a classification problem.
In this paper, we investigate this important design decision and also address complementary parameter choices such as the input representation. We evaluate a state-of-the-art DNN audio model based on a Bi-directional Long Short-Term Memory network architecture for speaker count estimations. Through experimental evaluations aimed at identifying the best overall strategy for the task and show results for five seconds speech segments in mixtures of up to ten speakers.
\end{abstract}
\section{Introduction}%
\label{sec:introduction}

In a ``cocktail-party'' scenario with many concurrent speakers, different applications may be envisioned such as localization, crowd monitoring, surveillance, speech recognition or speaker separation.
In this scenario, a typical assumption is that the number of concurrent speakers is known, which turns out to be of paramount importance for the effectiveness of subsequent processings.
Unfortunately, in real world applications, information about the actual number of concurrent speakers is often not available.
Surprisingly, very few methods have been proposed to address the task of counting the number of speakers.
\par

\begin{figure}[t]
  \centering
  \includegraphics[width=0.66\textwidth]{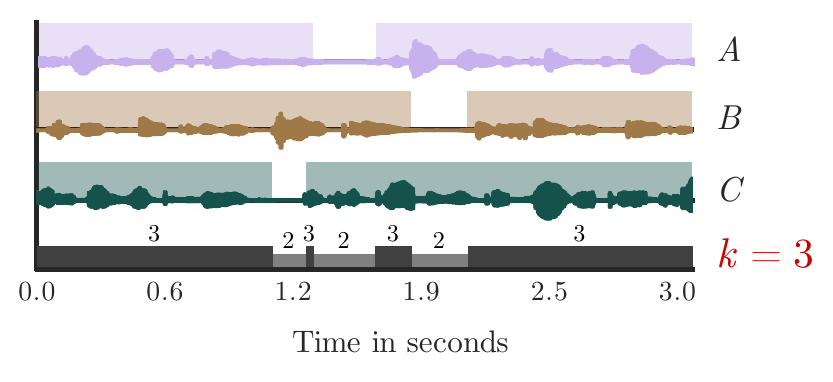}
  \caption{Illustration of our application scenario of three concurrent speakers and their respective speech activity. Bottom plot shows the number of concurrently active speakers and its maximum \(k\) which is our targeted output.}%
\label{fig:teaser}
\end{figure}

Estimating the maximum number of concurrent speakers is closely related to the more difficult problem of \textit{identifying} them, which is the topic of speaker diarisation (who speaks when)~\cite{tranter06}.
We call a system that identifies speakers first, ``counting by detection''.
These systems often use segments where only one speaker is active to discriminate the speakers.
Then comparisons of found segments are made to discriminate and temporally locate the speakers within a given recording.
When sources are fully overlapped as in real ``cocktail party'' environments, such a segmentation is hardly feasible.
And when a speaker overlap is as prevalent as in a ``cocktail-party'' scenario, developing an algorithm to detect speakers is challenging.
In this study we therefore attempt to directly estimate a speaker count instead of counting them after identification.
We refer to this strategy as ``direct count estimation''.

In multichannel signal processing, count estimation usually is achieved by estimating directions of arrival (DoA) and clustering them~\cite{mirzaei15, walter15}.
The first single channel method, based on thresholding amplitude modulation patterns, was proposed in~\cite{arai03}.
In~\cite{sayoud10}, the authors propose an energy feature based on temporally averaged mel-scale filter outputs.
In a more recent work~\cite{xu13}, the number of speakers is estimated by applying hierarchical clustering on fixed-length audio segments.
The main weakness of this method is to rely on the assumption that there are segments where only one speaker is active.
In another vein, Andrei et.al.~\cite{andrei15_interspeech} proposed an algorithm which correlates single frames of multi-speaker mixtures with a set of single-speaker utterances.
Motivated by the recent and impressive successes of deep learning approaches in various audio-related tasks~\cite{yu16, hershey16}, we focus on developing such a method for direct count estimation.
In computer vision, (object) count estimation using DNNs has recently achieved state-of-the-art performance~\cite{wang15, chattopadhyay17, khan16, segui15, zhang15, arteta16, marsden16, boominathan16, zhang2015salient}.
Two main paradigms may be found in the literature for this purpose, namely, \emph{regression} and \emph{classification}.

In this work we want to build upon these findings to achieve direct count estimation of speakers. Our main contributions are: i) to formulate the speaker count estimation problem as either a classification or a regression task,
and ii) to propose a neural network architecture based on a state-of-the-art BLSTM network, to infer the number of speakers from short audio segments of 5s.
Finally, we present experimental results for the different problem formulations as well as input feature representations to identify the best strategy for this task. For the sake of reproducibility, pre-trained network and the test dataset are made available for download on the accompanying website.\footnote{\url{https://www.audiolabs-erlangen.de/resources/2017-CountNet}}

\section{Problem Formulation}%
\label{sec:problem_formulation}
We consider the task of estimating the maximum number of concurrent speakers, \( \cardinality \in \mathbb{Z}^{+}_{0} \), in a single channel audio mixture \(\mathbf{x}\).
This is achieved by applying a mapping from \(\mathbf{x}\) to \(\cardinality \).
Let \(\mathbf{x}\) be a time domain signal of \(N\) samples, representing a linear mixture of \(L\) unique single speaker speech signals \(\mathbf{s}_l\).
Naturally, not all speakers~$l=1,\dots,L$ are active at every time instance.
We therefore, for each sample \(n\), introduce a latent binary \textit{speech activity} variable~$v_{nl}\in \left\{ 0,1 \right\}$.
Then, our task is to estimate
\begin{equation}
k=\underset{n}{\max}\left(\sum_{l=1}^{L}v_{nl}\right).
\label{eq:definition_k}
\end{equation}
It can be seen that our proposed task of estimating $k\leq L$ is more closely related to source separation whereas the estimation of \(L\) itself is more useful for tasks where speakers do not overlap.
We assume that no additional prior information except the maximum number of concurrent speakers, $k_{\max}$, is available, representing an upper limit for estimation.
In Figure~\ref{fig:teaser}, we illustrate our setup in a ``cocktail-party'' scenario featuring~$L=k=3$ speakers.
For the DNN system proposed in this paper, we use a non-negative time-frequency (TF) input representation \(\mathbf{X} \in\mathbb{R}_{+}^{D \times F}\) instead of \(\mathbf{x}\), where $D$ and $F$ denote the total number of time frames and frequency sub-bands, respectively.

\subsection{Estimation in a Deep Learning Framework}%
\label{ssec:estimation_framework}
In this study, we choose a deep neural network (DNN) as the mapping function $f_{\mathbf{\theta}}$ from the input~$\mathbf{X}$ to the output ~$y$, given by \(y=f_{\mathbf{\theta}}\left(\mathbf{X}\right)\),
where the optimal parameters (weights) $\mathbf{\theta}$ are learned via supervised training.
The output of the DNN is not necessarily the direct source count~$k$, therefore we introduce $q\left(\cdot\right)$ as a~\textit{decision function}, such that
\begin{equation}
\hat{k}=q\left(f_{\mathbf{\theta}}\left(\boldsymbol{X}\right)\right).
\end{equation}
The DNN is trained in a supervised manner using a training database of~$ \left\{\mathbf{X},k\right\}$ examples.
In this work, we want to investigate three different choices for the output distributions of the DNN, as well as the corresponding decision functions~$q\left(\cdot\right)$.

\textbf{Classification}: Here the output distribution is directly taken as \textit{discrete}, discarding any meaning concerning the ordering of the different possible values.
Given some particular input~$\mathbf{X}$, the network generates the posterior output probability for \((k_{\max} + 1)\) classes (including \(k=0\)) using the softmax activation function, and a maximum a posteriori (MAP) decision function is chosen that simply picks the most likely class \(q = \arg\max(\cdot)\).
Notwithstanding its conceptual simplicity, classification has two drawbacks.
First, the intuitive ranking between different estimates is lost: e.g. \(p(\cardinality = 6) \) may not depend on \(p(\cardinality = 5) \).
Second, the largest possible count $k_{\max}$ is given \textit{a priori}.
Despite these limitations, classification-based approaches have successfully been applied in deep neural networks for counting objects~\cite{segui15, zhang2015salient, khan16} in images.\\
\textbf{Gaussian Regression}: In regression, $k$ is derived from an output distribution defined on the real line. %
The output distribution in this setting is assumed to be Gaussian and the associated cost function is the classical squared error.
During inference and given the output~$f_{\mathbf{\mathbf{\theta}}}\left(\mathbf{X}\right)$ of the network, the best discrete value that is consistent with the model is simply the rounding operator $q = \left[\cdot\right]$.
Gaussian regression has achieved state-of-the-art counting performance in computer vision using deep learning frameworks~\cite{zhang15, arteta16, marsden16, boominathan16}.

\textbf{Discrete Poisson modelling}: When it comes to modelling count data, it is often shown effective to adopt the Poisson distribution~\cite{fallah09}.
First, this strategy retains the advantage of the classification approach to directly pick a probabilistic model over the actual discrete observations, avoiding the somewhat artificial trick of introducing a latent variable that would be rounded to yield the observation.
Second, the model avoids the inconvenience of the classification approach to completely drop dependencies between classes.
Due to these advantages, the Poisson distribution has been used in studies devising deep architectures for counting systems~\cite{Rezatofigh16}.
For instance in~\cite{fallah09, chan09, Rezatofigh16}, it is shown that the number of objects in images can be well modelled by the Poisson distribution. Inspired by these previous works, we also consider the Poisson output distribution \(\mathcal{P}\left(k\mid f_{\mathbf{\mathbf{\theta}}}\left(\mathbf{X}\right)\right)\)
where $\mathcal{P}\left(\cdot\mid\lambda\right)$ denotes the Poisson distribution with scale parameter~$\lambda$.
In this setup, the cost function at learning time is the Poisson negative log-likelihood and the deep architecture at test time provides the predicted scale parameter $f_{\mathbf{\mathbf{\theta}}}\left(\mathbf{X}\right) \in\mathbb{R}_+$, which summarizes the whole output distribution.
As a decision function~$q$ in this setting, we considered several alternatives. A first option is to again resort to MAP estimation and pick the mode $\left[f_{\mathbf{\mathbf{\theta}}}\left(\mathbf{X}\right)\right]$ of the distribution as a point estimate. However, experiments showed that the posterior median yields better estimates, and is given by
\begin{subequations}
\begin{align}
q\left(f_{\mathbf{\mathbf{\theta}}}\left(\mathbf{X}\right)\right) & =\underset{\hat{k}}{\text{argmin}}\sum_{k=0}^{\infty}\left|\hat{k}-k\right|\mathcal{P}\left(k\mid f_{\mathbf{\mathbf{\theta}}}\left(\mathbf{X}\right)\right)\\
 & =\text{median}\left(k\sim\mathcal{P}\left(f_{\mathbf{\mathbf{\theta}}}\left(\mathbf{X}\right)\right)\right)
 \label{eq:decision_function_poisson}
\end{align}
\end{subequations}
where the median of a Poisson distributed random variable was approximated given the expression in~\cite{Choi94}.

\section{Proposed Model}%
\label{sec:supervised_learning}
Various audio-related applications share common DNN architecture designs, often found by incorporating domain knowledge and through extensive hyperparameter searches.
For our proposed task of source count estimation, however, domain knowledge is difficult to incorporate, as this study aims at revealing the best strategy to address the problem.
Therefore, we use a network built upon an existing BLSTM-RNN architecture, that has already shown a considerable amount of generalization for various audio applications~\cite{Leglaive15, hagerer17}.

A recurrent neural network (RNN) is very similar to a fully connected network, except that RNN applies the same set of weights recursively over an input sequence.
RNNs can detect structure in sequential data of arbitrary length. %
This makes it ideal to model time series, however, in practice, the temporal context learnt is limited to only a few time instances, because of the vanishing gradient problem~\cite{Hochreiter98}.

To alleviate this problem, forgetting factors (also called gating) were proposed.
One of the most popular gated recurrent cells is the Long Short-Term Memory (LSTM)~\cite{Hochreiter97} cell.
Its effectiveness has been proven in various applications and LSTMs are the state-of-the-art approach for speech recognition~\cite{Graves13} or singing voice detection~\cite{Leglaive15}.
In this work we employed a bi-directional LSTM (BLSTM) with three hidden layers whose sizes are 30, 20 and 40 similar to the architecture introduced in~\cite{Leglaive15}. A BLSTM is more robust compared to a simple LSTM, since input information from both past as well as the future in used to learn the weights. For further information on BLSTMs, the reader is referred to~\cite{Goodfellow16}.
\par
For a given input sequence,
the output of a recurrent layer is either only the last step output or a full sequence.
We found that employing the full sequence output of the last recurrent layer before feeding it into the fully connected output layer is important in the context of RNNs for count estimation.
Furthermore we added a temporal max pooling layer with pooling size 2 to reduce the number of parameters for the fully connected layer.
Temporal max pooling intuitively fits to our problem formulation which in itself is a maximum of the number of sources in a specific number of frames.

As we introduced in Section~\ref{ssec:estimation_framework}, the count estimation problem can be addressed using three different strategies.
For each of the decision functions a suitable output activation and loss is used as shown in Table~\ref{tab:outputs}. Except for these (output) parameters, all models have the same parameters.

\begin{table}
  \centering
\begin{tabular}{llll}
\toprule
Output Type & Activation & Dim. & Loss \\
\midrule
Classification & Softmax & \(\mathbb{B}^{k_{\max} + 1}\) & Cat. cross entropy \\
Gaussian Regr. & Linear & \(\mathbb{R}^{1}\) & MSE \\
Poisson Regr. & Exponential & \(\mathbb{R}^{1}\) & Neg. log likelihood\\
\bottomrule
\end{tabular}
\caption{Output Activation Functions and Loss Functions}%
\label{tab:outputs}
\end{table}
\section{Training}%
\label{sec:training}

Since a realistic dataset of fully overlapped speakers is not available, we  chose to generate synthetic mixtures.
We recognize that in a simulated ``cocktail-party'' environment, mixtures lack  the conversational aspect of human communication but provide a controlled environment which helps understand how a DNN solves the count estimation problem.
As we aim for a speaker independent solution, we selected a speech corpus with a large number of different speakers instead of large number of utterances, yielding a larger number of unique mixtures.
For training we selected the \emph{LibriSpeech clean-360} \cite{panayotov15} dataset which includes 363 hours of clean speech of English utterances from 921 speakers (439 female and 482 male speakers).

As revealed in Section~\ref{sec:problem_formulation}, the maximum number of concurrent speakers \( \cardinality \) requires annotation of the activity of each individual speaker.
Even though \emph{LibriSpeech} comes with annotations, they often are not consistent across different corpora.
We therefore generated annotations based on a voice activity detection algorithm (VAD). In this work, we used the implementation from the \emph{Chromium Web Browser} that is part of the WebRTC Standard~\cite{webrtc}.

\par

To generate a single training sample \(\left\{\mathbf{X},k\right\}\), we draw a unique set of \(L\) speakers from the corpus.
For each of the speakers we then select a random utterance, resampled to 16 kHz sampling rate and apply VAD.\@
The VAD method was configured using a hop size of 10~ms.
Further, the VAD estimate was used to remove silence in the beginning and the end of an utterance recording.
In the next step, more utterances from the same speaker are drawn from the corpus until the desired duration is reached.
Both, the audio recording and the VAD annotation of each utterance is concatenated.
The procedure is repeated for all speakers so that \(L\) time domain signals are created.
The signals are mixed and peak normalized to avoid clipping.
Mixtures are then transformed to a time-frequency matrix \(X \in D \times F\) as defined in Section~\ref{sec:problem_formulation}.
The ground truth output \(\cardinality\) are then computed using the VAD matrix based on Equation~\ref{eq:definition_k}.
\par

We follow the proposal of~\cite{wang15} and include non-speech examples in our training data to avoid using zero input samples for \(\cardinality = 0\).
For this, we used the TUT Acoustic Scenes dataset~\cite{Mesaros16} to create negative training samples using the same procedure as described above.
Because these environmental sounds could include speech, scenes with \texttt{cafe/restaurant}, \texttt{grocery store} and \texttt{metro station} were omitted.

As our application closely relates to source separation it is desirable for our trained DNN system to be robust against gain variations.
We therefore find it important to make sure that the DNN cannot leverage the gain factors of the mixture.
We found that the averaged energy of one bin across all frames of the input sample already is a solid indicator for the number of speakers.

To accomodate these findings, we normalize \(\mathbf{X}\) to the average Euclidean norm of all frames as used in~\cite{Uhlich15}.
Additionally, as common in machine learning, we scale the normalized input representation so that the feature dimensions have zero mean and unit variance/standard deviation across the whole training dataset.
\par
To train the network we use Poisson sampling to balance the number of samples \(T_{ \cardinality}\) for each \(\cardinality \).
For our experiments we chose a medium-sized training dataset of \(T_{\cardinality}\ = 1820 \; \text{samples} \; \forall \cardinality \in [0, \ldots, 10]\) resulting in a total of 20,020 training items, each containing 10 seconds of audio, resulting in 55.55 hours of training material.
The actual duration of each input is reduced to five seconds by selecting a random excerpt from each mixture. For each excerpt Equation~\ref{eq:definition_k} is evaluated to generate a single sample, then combined into mini-batches of 32 samples.
This way the network is seeing slightly different samples (in different order) in each training epoch.
We found this procedure (also used in~\cite{schluter16}) to help speeding up the stochastic gradient based training process.
The DNN is trained using the ADAM optimizer~\cite{kingma14}. %
In addition to the training dataset we created a separate validation dataset of \(T_{\cardinality}\ = 5720\) samples using a different set of speakers from \emph{LibriSpeech dev-clean}.
Early stopping is applied by monitoring the validation loss to reduce the effect of overfitting.
Training never exceeded 50 epochs.
\vspace{-0.2cm}
\section{Evaluation}%
\label{sec:evaluation}

We evaluated our proposed network architecture with two main parameters: the three proposed output distributions (see Section~\ref{sec:problem_formulation}) and four different input representations.
To allow for a controlled test environment and at the same time limit the number of training iterations, we fix certain parameters:
In our experiment all speakers were mixed to 0~dB SNR.\@
For all experimental parameters we ran the training three times with different random seeds for each run and report averaged the results to minimize random effects caused by early stopping.
We used the \emph{LibriSpeech test-clean} subsets to generate 5720 unique and unseen speaker mixtures of five seconds duration for the test set with
\(k_{\max} = L = 10\).
\begin{figure*}[h!]
  \centering
  \begin{minipage}[b]{\textwidth}
  \centering
  \begin{subfigure}[b]{0.49\textwidth}
      \begin{adjustbox}{width=\textwidth}
\begin{tikzpicture}

\definecolor{color3}{rgb}{0.9450980392,0.6392156863,0.2509803922}
\definecolor{color0}{rgb}{0.927647058823529,0.927647058823529,0.927647058823529}
\definecolor{color1}{rgb}{0.36862745,  0.23529412,  0.6}
\definecolor{color2}{rgb}{0.59803922,  0.7745098 ,  0.44117647}

\begin{axis}[
xlabel={$k$},
ylabel={MAE},
xmin=-0.5, xmax=10.5,
ymin=-0.0792013957712289, ymax=1.33125716523073,
xtick={0,1,2,3,4,5,6,7,8,9,10},
xticklabels={0,1,2,3,4,5,6,7,8,9,10},
ytick={-0.2,0,0.2,0.4,0.6,0.8,1,1.2,1.4},
yticklabels={0.0,0.0,0.2,0.4,0.6,0.8,1.0,1.2,},
tick align=outside,
tick pos=left,
x grid style={white},
ymajorgrids,
xmajorgrids,
width=\textwidth,
height=0.9\textwidth,
y grid style={white},
axis line style={white},
axis background/.style={fill=color0},
legend style={{font=\fontsize{7}{8}\selectfont}, at={(0.03,0.97)}, anchor=north west, draw=none, fill=color0},
legend entries={{Classification},{Gaussian Regr.},{Poisson Regr.}},
legend cell align={left}
]
\addplot [only marks, draw=color1, fill=color1, colormap/blackwhite]
table{%
x                      y
-3.750000000000001e-02 +6.410256410256410e-04
+9.625000000000000e-01 +2.259332023575639e-02
+1.962500000000000e+00 +1.280684754521964e-01
+2.962500000000000e+00 +2.663427561837455e-01
+3.962500000000000e+00 +3.709625322997416e-01
+4.962500000000000e+00 +5.164431673052363e-01
+5.962500000000000e+00 +5.981404958677685e-01
+6.962500000000000e+00 +7.687969924812030e-01
+7.962500000000000e+00 +8.047900262467192e-01
+8.962500000000000e+00 +8.131313131313130e-01
+9.962500000000000e+00 +4.267676767676769e-01
};
\addplot [line width=1.00pt, color1, forget plot]
table {%
-0.0375 0.000641025641025641
0.9625 0.0225933202357564
1.9625 0.128068475452196
2.9625 0.266342756183746
3.9625 0.370962532299742
4.9625 0.516443167305236
5.9625 0.598140495867769
6.9625 0.768796992481203
7.9625 0.804790026246719
8.9625 0.813131313131313
9.9625 0.426767676767677
};
\addplot [line width=1.00pt, color1, forget plot]
table {%
-0.0375 0
-0.0375 0.0016025641025641
};
\addplot [line width=1.00pt, color1, forget plot]
table {%
0.9625 0.0145628683693517
0.9625 0.0320890635232482
};
\addplot [line width=1.00pt, color1, forget plot]
table {%
1.9625 0.0960876937984496
1.9625 0.164256298449612
};
\addplot [line width=1.00pt, color1, forget plot]
table {%
2.9625 0.2262882803298
2.9625 0.309190959952886
};
\addplot [line width=1.00pt, color1, forget plot]
table {%
3.9625 0.316210432816537
3.9625 0.419101259689923
};
\addplot [line width=1.00pt, color1, forget plot]
table {%
4.9625 0.430707215836526
4.9625 0.600139687100894
};
\addplot [line width=1.00pt, color1, forget plot]
table {%
5.9625 0.501291322314049
5.9625 0.696121728650138
};
\addplot [line width=1.00pt, color1, forget plot]
table {%
6.9625 0.69046835839599
6.9625 0.842575187969925
};
\addplot [line width=1.00pt, color1, forget plot]
table {%
7.9625 0.714554625984252
7.9625 0.891437007874016
};
\addplot [line width=1.00pt, color1, forget plot]
table {%
8.9625 0.722516835016835
8.9625 0.897988215488216
};
\addplot [line width=1.00pt, color1, forget plot]
table {%
9.9625 0.292589962121212
9.9625 0.610483743686869
};
\addplot [only marks, draw=color2, fill=color2, colormap/blackwhite]
table{%
x                      y
+0.000000000000000e+00 +9.615384721352408e-04
+1.000000000000000e+00 +4.780615648875634e-02
+2.000000000000000e+00 +1.889534884442886e-01
+3.000000000000000e+00 +3.303886937598388e-01
+4.000000000000000e+00 +4.691537419954936e-01
+5.000000000000000e+00 +6.198914547761282e-01
+6.000000000000000e+00 +7.358815521001816e-01
+7.000000000000000e+00 +7.742794503768285e-01
+8.000000000000000e+00 +6.369750648736954e-01
+9.000000000000000e+00 +5.127946138381958e-01
+1.000000000000000e+01 +6.071654036641121e-01
};
\addplot [line width=1.00pt, color2, forget plot]
table {%
0 0.000961538472135241
1 0.0478061564887563
2 0.188953488444289
3 0.330388693759839
4 0.469153741995494
5 0.619891454776128
6 0.735881552100182
7 0.774279450376829
8 0.636975064873695
9 0.512794613838196
10 0.607165403664112
};
\addplot [line width=1.00pt, color2, forget plot]
table {%
0 0.000160256410405661
0 0.00192307695397176
};
\addplot [line width=1.00pt, color2, forget plot]
table {%
1 0.0265185005535992
1 0.0736738717610327
};
\addplot [line width=1.00pt, color2, forget plot]
table {%
2 0.153088661360865
2 0.226102230853091
};
\addplot [line width=1.00pt, color2, forget plot]
table {%
3 0.276060073326031
3 0.397232040762901
};
\addplot [line width=1.00pt, color2, forget plot]
table {%
4 0.396455096205076
4 0.575637913805743
};
\addplot [line width=1.00pt, color2, forget plot]
table {%
5 0.521220470344027
5 0.73771154868106
};
\addplot [line width=1.00pt, color2, forget plot]
table {%
6 0.63892045840621
6 0.845742959529161
};
\addplot [line width=1.00pt, color2, forget plot]
table {%
7 0.677001087119182
7 0.89132988974452
};
\addplot [line width=1.00pt, color2, forget plot]
table {%
8 0.573978828266263
8 0.712955214455724
};
\addplot [line width=1.00pt, color2, forget plot]
table {%
9 0.469507576711476
9 0.557596805940072
};
\addplot [line width=1.00pt, color2, forget plot]
table {%
10 0.486576702507834
10 0.74371054880321
};

\addplot [only marks, draw=color3, fill=color3, colormap/blackwhite]
table{%
x                      y
+3.750000000000001e-02 +1.121794871794872e-03
+1.037500000000000e+00 +1.113294040602489e-02
+2.037500000000000e+00 +1.732881136950905e-01
+3.037500000000000e+00 +3.194935217903416e-01
+4.037500000000000e+00 +4.211886304909561e-01
+5.037500000000000e+00 +5.416666666666666e-01
+6.037500000000000e+00 +6.236225895316805e-01
+7.037500000000000e+00 +6.475563909774437e-01
+8.037500000000000e+00 +6.478018372703412e-01
+9.037500000000000e+00 +6.912457912457913e-01
+1.003750000000000e+01 +8.803661616161618e-01
};
\addplot [line width=1.00pt, color3, forget plot]
table {%
0.0375 0.00112179487179487
1.0375 0.0111329404060249
2.0375 0.17328811369509
3.0375 0.319493521790342
4.0375 0.421188630490956
5.0375 0.541666666666667
6.0375 0.62362258953168
7.0375 0.647556390977444
8.0375 0.647801837270341
9.0375 0.691245791245791
10.0375 0.880366161616162
};
\addplot [line width=1.00pt, color3, forget plot]
table {%
0.0375 0
0.0375 0.00272435897435897
};
\addplot [line width=1.00pt, color3, forget plot]
table {%
1.0375 0.00605762933857236
1.0375 0.0173583824492469
};
\addplot [line width=1.00pt, color3, forget plot]
table {%
2.0375 0.135493378552972
2.0375 0.215128391472868
};
\addplot [line width=1.00pt, color3, forget plot]
table {%
3.0375 0.283119846878681
3.0375 0.36588265606596
};
\addplot [line width=1.00pt, color3, forget plot]
table {%
4.0375 0.397771317829457
4.0375 0.443802487080103
};
\addplot [line width=1.00pt, color3, forget plot]
table {%
5.0375 0.514683109833972
5.0375 0.571208492975734
};
\addplot [line width=1.00pt, color3, forget plot]
table {%
6.0375 0.588304924242424
6.0375 0.657558539944904
};
\addplot [line width=1.00pt, color3, forget plot]
table {%
7.0375 0.600716635338346
7.0375 0.692046522556391
};
\addplot [line width=1.00pt, color3, forget plot]
table {%
8.0375 0.596604330708662
8.0375 0.718836122047244
};
\addplot [line width=1.00pt, color3, forget plot]
table {%
9.0375 0.5765867003367
9.0375 0.829124579124579
};
\addplot [line width=1.00pt, color3, forget plot]
table {%
10.0375 0.678484059343434
10.0375 1.1229521780303
};
\end{axis}

\end{tikzpicture}
       \end{adjustbox}
      \caption{Output Distribution}%
      \label{fig:ssec:exp_fixed_gains/A}
  \end{subfigure}
  \begin{subfigure}[b]{0.49\textwidth}
      \begin{adjustbox}{width=\textwidth}
\begin{tikzpicture}

\definecolor{color1}{rgb}{0,0.447058823529412,0.698039215686274}
\definecolor{color0}{rgb}{0.927647058823529,0.927647058823529,0.927647058823529}
\definecolor{color3}{rgb}{0.835294117647059,0.368627450980392,0}
\definecolor{color2}{rgb}{0,0.619607843137255,0.450980392156863}
\definecolor{color4}{rgb}{0.8,0.474509803921569,0.654901960784314}

\begin{axis}[
xlabel={$k$},
ylabel={MAE},
xmin=-0.5, xmax=10.5,
ymin=-0.0772346843650762, ymax=1.27547568471466,
xtick={0,1,2,3,4,5,6,7,8,9,10},
xticklabels={0,1,2,3,4,5,6,7,8,9,10},
ytick={-0.2,0,0.2,0.4,0.6,0.8,1,1.2,1.4},
yticklabels={0.0,0.0,0.2,0.4,0.6,0.8,1.0,1.2,},
tick align=outside,
tick pos=left,
x grid style={white},
ymajorgrids,
xmajorgrids,
width=\textwidth,
height=0.9\textwidth,
y grid style={white},
axis line style={white},
axis background/.style={fill=color0},
legend style={{font=\fontsize{7}{8}\selectfont}, at={(0.03,0.97)}, anchor=north west, draw=none, fill=color0},
legend cell align={left},
legend entries={{MEL40},{MFCC20},{STFT},{STFTLOG}},
]
\addplot [only marks, draw=color1, fill=color1, colormap/blackwhite]
table{%
x                      y
-5.000000000000000e-02 +0.000000000000000e+00
+9.500000000000000e-01 +4.933420704265813e-02
+1.950000000000000e+00 +2.204995712876628e-01
+2.950000000000000e+00 +3.788771103375745e-01
+3.950000000000000e+00 +5.094745938989440e-01
+4.950000000000000e+00 +6.511281488701961e-01
+5.950000000000000e+00 +7.334710815502373e-01
+6.950000000000000e+00 +8.051378526783228e-01
+7.950000000000000e+00 +8.011811009542211e-01
+8.949999999999999e+00 +8.143658785038659e-01
+9.949999999999999e+00 +9.071969720048936e-01
};
\addplot [line width=1.00pt, color1, forget plot]
table {%
-0.05 0
0.95 0.0493342070426581
1.95 0.220499571287663
2.95 0.378877110337575
3.95 0.509474593898944
4.95 0.651128148870196
5.95 0.733471081550237
6.95 0.805137852678323
7.95 0.801181100954221
8.95 0.814365878503866
9.95 0.907196972004894
};
\addplot [line width=1.00pt, color1, forget plot]
table {%
-0.05 0
-0.05 0
};
\addplot [line width=1.00pt, color1, forget plot]
table {%
0.95 0.0255348177253875
0.95 0.0796878426936913
};
\addplot [line width=1.00pt, color1, forget plot]
table {%
1.95 0.169891257357241
1.95 0.269616714828069
};
\addplot [line width=1.00pt, color1, forget plot]
table {%
2.95 0.31468394176693
2.95 0.448208677717223
};
\addplot [line width=1.00pt, color1, forget plot]
table {%
3.95 0.431734496167669
3.95 0.619745913671585
};
\addplot [line width=1.00pt, color1, forget plot]
table {%
4.95 0.538516391462318
4.95 0.815948297769123
};
\addplot [line width=1.00pt, color1, forget plot]
table {%
5.95 0.610192840694834
5.95 0.86939280335924
};
\addplot [line width=1.00pt, color1, forget plot]
table {%
6.95 0.687969929147932
6.95 0.931302230999982
};
\addplot [line width=1.00pt, color1, forget plot]
table {%
7.95 0.710192477046035
7.95 0.901607611985292
};
\addplot [line width=1.00pt, color1, forget plot]
table {%
8.95 0.678013464906467
8.95 0.982716045421449
};
\addplot [line width=1.00pt, color1, forget plot]
table {%
9.95 0.594470750517917
9.95 1.23772622229977
};
\addplot [only marks, draw=color2, fill=color2, colormap/blackwhite]
table{%
x                      y
-1.666666666666667e-02 +0.000000000000000e+00
+9.833333333333333e-01 +3.318052821770502e-02
+1.983333333333333e+00 +1.627906959930645e-01
+2.983333333333333e+00 +2.872006293835245e-01
+3.983333333333333e+00 +4.000861291018773e-01
+4.983333333333333e+00 +5.721583698038448e-01
+5.983333333333333e+00 +6.657483943894547e-01
+6.983333333333333e+00 +7.504177047694437e-01
+7.983333333333333e+00 +6.942257231406131e-01
+8.983333333333333e+00 +5.842873207112621e-01
+9.983333333333333e+00 +4.313973045911050e-01
};
\addplot [line width=1.00pt, color2, forget plot]
table {%
-0.0166666666666667 0
0.983333333333333 0.033180528217705
1.98333333333333 0.162790695993065
2.98333333333333 0.287200629383525
3.98333333333333 0.400086129101877
4.98333333333333 0.572158369803845
5.98333333333333 0.665748394389455
6.98333333333333 0.750417704769444
7.98333333333333 0.694225723140613
8.98333333333333 0.584287320711262
9.98333333333333 0.431397304591105
};
\addplot [line width=1.00pt, color2, forget plot]
table {%
-0.0166666666666667 0
-0.0166666666666667 0
};
\addplot [line width=1.00pt, color2, forget plot]
table {%
0.983333333333333 0.0176544422399854
0.983333333333333 0.0517354284887664
};
\addplot [line width=1.00pt, color2, forget plot]
table {%
1.98333333333333 0.119939706790278
1.98333333333333 0.213393619020921
};
\addplot [line width=1.00pt, color2, forget plot]
table {%
2.98333333333333 0.24457204693034
2.98333333333333 0.337666866377517
};
\addplot [line width=1.00pt, color2, forget plot]
table {%
3.98333333333333 0.333758609516676
3.98333333333333 0.465127037912222
};
\addplot [line width=1.00pt, color2, forget plot]
table {%
4.98333333333333 0.519359312023493
4.98333333333333 0.632827803652405
};
\addplot [line width=1.00pt, color2, forget plot]
table {%
5.98333333333333 0.548789027724603
5.98333333333333 0.771814738880395
};
\addplot [line width=1.00pt, color2, forget plot]
table {%
6.98333333333333 0.68901420088614
6.98333333333333 0.827730780956938
};
\addplot [line width=1.00pt, color2, forget plot]
table {%
7.98333333333333 0.625967855918387
7.98333333333333 0.771872264099872
};
\addplot [line width=1.00pt, color2, forget plot]
table {%
8.98333333333333 0.516049388406587
8.98333333333333 0.671391695267557
};
\addplot [line width=1.00pt, color2, forget plot]
table {%
9.98333333333333 0.308059764284678
9.98333333333333 0.543386986074022
};

\addplot [only marks, draw=color3, fill=color3, colormap/blackwhite]
table{%
x                      y
+1.666666666666666e-02 +1.068376068774070e-03
+1.016666666666667e+00 +1.178781938954824e-02
+2.016666666666667e+00 +1.307062879268924e-01
+3.016666666666667e+00 +2.638398129646573e-01
+4.016666666666667e+00 +3.548664931435712e-01
+5.016666666666667e+00 +4.763729270213003e-01
+6.016666666666667e+00 +5.445362752572017e-01
+7.016666666666667e+00 +6.263575650994342e-01
+8.016666666666667e+00 +6.170166188546262e-01
+9.016666666666667e+00 +6.473625136010441e-01
+1.001666666666667e+01 +6.632996624969072e-01
};
\addplot [line width=1.00pt, color3, forget plot]
table {%
0.0166666666666667 0.00106837606877407
1.01666666666667 0.0117878193895482
2.01666666666667 0.130706287926892
3.01666666666667 0.263839812964657
4.01666666666667 0.354866493143571
5.01666666666667 0.4763729270213
6.01666666666667 0.544536275257202
7.01666666666667 0.626357565099434
8.01666666666667 0.617016618854626
9.01666666666667 0.647362513601044
10.0166666666667 0.663299662496907
};
\addplot [line width=1.00pt, color3, forget plot]
table {%
0.0166666666666667 0.000213675213874214
0.0166666666666667 0.00192307692387293
};
\addplot [line width=1.00pt, color3, forget plot]
table {%
1.01666666666667 0.00742196035877619
1.01666666666667 0.0165902644174179
};
\addplot [line width=1.00pt, color3, forget plot]
table {%
2.01666666666667 0.103138457616016
2.01666666666667 0.157412791026886
};
\addplot [line width=1.00pt, color3, forget plot]
table {%
3.01666666666667 0.226732429847689
3.01666666666667 0.300942286036245
};
\addplot [line width=1.00pt, color3, forget plot]
table {%
4.01666666666667 0.325355296669902
4.01666666666667 0.379414295554033
};
\addplot [line width=1.00pt, color3, forget plot]
table {%
5.01666666666667 0.40547573544361
5.01666666666667 0.544082590648662
};
\addplot [line width=1.00pt, color3, forget plot]
table {%
6.01666666666667 0.501825070798507
6.01666666666667 0.588842979237725
};
\addplot [line width=1.00pt, color3, forget plot]
table {%
7.01666666666667 0.562447790618528
7.01666666666667 0.691765875866638
};
\addplot [line width=1.00pt, color3, forget plot]
table {%
8.01666666666667 0.551175626361464
8.01666666666667 0.699278211609272
};
\addplot [line width=1.00pt, color3, forget plot]
table {%
9.01666666666667 0.528597081632392
9.01666666666667 0.773540965222851
};
\addplot [line width=1.00pt, color3, forget plot]
table {%
10.0166666666667 0.469460226404747
10.0166666666667 0.902598903792274
};
\addplot [only marks, draw=color4, fill=color4, colormap/blackwhite]
table{%
x                      y
+5.000000000000000e-02 +2.564102577833602e-03
+1.050000000000000e+00 +1.440733485747209e-02
+2.050000000000000e+00 +1.397502149144809e-01
+3.050000000000000e+00 +2.917157429594782e-01
+4.050000000000000e+00 +4.173126569038625e-01
+5.050000000000000e+00 +5.376756059687002e-01
+6.050000000000000e+00 +6.664370988026138e-01
+7.050000000000000e+00 +7.389306559000995e-01
+8.050000000000001e+00 +6.736657949048808e-01
+9.050000000000001e+00 +6.435465781375616e-01
+1.005000000000000e+01 +5.505050503043614e-01
};
\addplot [line width=1.00pt, color4, forget plot]
table {%
0.05 0.0025641025778336
1.05 0.0144073348574721
2.05 0.139750214914481
3.05 0.291715742959478
4.05 0.417312656903862
5.05 0.5376756059687
6.05 0.666437098802614
7.05 0.7389306559001
8.05 0.673665794904881
9.05 0.643546578137562
10.05 0.550505050304361
};
\addplot [line width=1.00pt, color4, forget plot]
table {%
0.05 0.000854700868426918
0.05 0.00448717951424854
};
\addplot [line width=1.00pt, color4, forget plot]
table {%
1.05 0.00785854626229684
1.05 0.0240176821835861
};
\addplot [line width=1.00pt, color4, forget plot]
table {%
2.05 0.102912359958086
2.05 0.174644701583287
};
\addplot [line width=1.00pt, color4, forget plot]
table {%
3.05 0.249293286086568
3.05 0.339031209053072
};
\addplot [line width=1.00pt, color4, forget plot]
table {%
4.05 0.377255595644212
4.05 0.462537673636822
};
\addplot [line width=1.00pt, color4, forget plot]
table {%
5.05 0.471237759145905
5.05 0.622844826140179
};
\addplot [line width=1.00pt, color4, forget plot]
table {%
6.05 0.592056937367956
6.05 0.748461892527654
};
\addplot [line width=1.00pt, color4, forget plot]
table {%
7.05 0.659121751329654
7.05 0.836680241890171
};
\addplot [line width=1.00pt, color4, forget plot]
table {%
8.05 0.573889991229079
8.05 0.780900045564481
};
\addplot [line width=1.00pt, color4, forget plot]
table {%
9.05 0.524562289451743
9.05 0.761470258544491
};
\addplot [line width=1.00pt, color4, forget plot]
table {%
10.05 0.382970326311058
10.05 0.708769997357319
};
\end{axis}

\end{tikzpicture}
       \end{adjustbox}
      \caption{Feature Representation}%
      \label{fig:ssec:exp_fixed_gains/B}
  \end{subfigure}
  \caption{Mean absolute error (MAE) over ground truth count \(k=[0\ldots10]\). Error bars show the 95\% confidence intervals. Results are averaged over the different feature representations (a) and output distributions (b).}
  \label{fig:ssec:exp_fixed_gains}

  \vspace{3ex}
  \end{minipage}\qquad
  \begin{minipage}[b]{0.5\textwidth}
  \centering
      \begin{adjustbox}{width=\textwidth}
\begin{tikzpicture}

\definecolor{color0}{rgb}{0.927647058823529,0.927647058823529,0.949019607843137}

\begin{axis}[
xlabel={$k$},
ylabel={$\hat{k}$},
xmin=0, xmax=11,
ymin=0, ymax=11,
xtick={0.5,1.5,2.5,3.5,4.5,5.5,6.5,7.5,8.5,9.5,10.5},
xticklabels={0,1,2,3,4,5,6,7,8,9,10},
ytick={0.5,1.5,2.5,3.5,4.5,5.5,6.5,7.5,8.5,9.5,10.5},
yticklabels={0,1,2,3,4,5,6,7,8,9,10},
tick align=outside,
tick pos=left,
width=1\textwidth,
height=1\textwidth,
font=\fontsize{11}{12}\selectfont,
x grid style={white},
y grid style={white},
axis line style={white},
axis background/.style={fill=color0},
colorbar,
colormap={mymap}{[1pt]
  rgb(0pt)=(0.968627450980392,0.988235294117647,0.992156862745098);
  rgb(1pt)=(0.898039215686275,0.96078431372549,0.976470588235294);
  rgb(2pt)=(0.8,0.925490196078431,0.901960784313726);
  rgb(3pt)=(0.6,0.847058823529412,0.788235294117647);
  rgb(4pt)=(0.4,0.76078431372549,0.643137254901961);
  rgb(5pt)=(0.254901960784314,0.682352941176471,0.462745098039216);
  rgb(6pt)=(0.137254901960784,0.545098039215686,0.270588235294118);
  rgb(7pt)=(0,0.427450980392157,0.172549019607843);
  rgb(8pt)=(0,0.266666666666667,0.105882352941176)
},
point meta min=0,
point meta max=1,
colorbar style={width=4pt, axis line style={draw=none}, ytick={0,0.2,0.4,0.6,0.8,1},yticklabels={0.0,0.2,0.4,0.6,0.8,1.0},ylabel={}}
]
\addplot graphics [includegraphics cmd=\pgfimage,xmin=0, xmax=11, ymin=0, ymax=11] {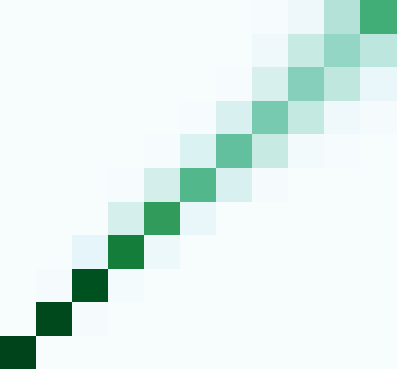};

\end{axis}

\end{tikzpicture}
       \end{adjustbox}
  \caption{Normalized confusion matrix showing $\hat{k}$ over $k$ for the test data of the best performing network (Output Distribution: Classification, Feature Representation: STFT).}
  \vspace{2ex}
 \label{fig:confusion}

  \end{minipage}%
\end{figure*}
\\
Since we are dealing with a novel task description, related speaker count estimation techniques like those introduced in Section~\ref{sec:introduction}, could hardly be used as baselines.
Specifically,~\cite{xu13} does not work on fully overlapped speech,~\cite{andrei15_interspeech} does not scale to the size of our dataset, since it requires to cross-correlate the full database against another.
Finally,~\cite{sayoud10} proposes a feature but does not employ a fully automated system. We translated this method into a data-driven approach and employed a vector quantizer to get an optimal mapping with respect to the sum of squares criterion (we refer to this as \emph{VQ}).

\subsection{Input Representations}

For our task, we chose several different input representations, well-established in speech application. We expect that a high frequency resolution is needed to discriminate time frequency bins with overlapped speech segments from those that only belong to a single speaker.
We compared the following input representations that were all based on a frame length of 25~ms:\\
\textbf{STFT}: magnitude of the Short-time Fourier transform computed using Hann windows.
The resulting input is \(X \in \mathbb{R}^{500 \times 201}\).\\
\textbf{LOGSTFT}: logarithmically scaled magnitudes from STFT representation using \(\log(1 + STFT)\).
The resulting input is \(X \in \mathbb{R}^{500 \times 201}\).\\
\textbf{MEL40}: compute mapping from the STFT output directly onto Mel basis using 40 triangular filters.
The resulting input is \(X \in \mathbb{R}^{500 \times 40}\).\\
\textbf{MFCC20}: First 20 Mel-frequency cepstral coefficients.
The resulting input is \(X \in \mathbb{R}^{500 \times 20}\).
\par
Before feature transformation, all input files were resampled to 16 kHz sampling rate. All features are computed using a hop size of 10~ms.

\subsection{Metric}%
\label{ssec:metric}

While the intermediate output \(y\) is treated as either a classification or a regression problem (See Section~\ref{sec:problem_formulation}), we evaluate the final output \(\cardinality \) as a discrete regression problem.
We therefore evaluate the performance using the mean absolute error (MAE), which is also used to evaluate other count related tasks (c.f.~\cite{zhang15, Rezatofigh16}).

\subsection{Results}
\label{ssec:results}
To find the best parameters, we performed training and evaluation for the parameters, resulting in 36 trained networks.
On average each network was trained 33 epochs before early stopping was engaged. Training duration was about 800 seconds per epoch on a NVIDIA GTX 1080 GPU.\@
We present the results in terms of input representation and output distribution in Figure~2.
One can see that the overall trend of the count error in MAE is similar regardless of the parametrisation: all models are able to reliably distinguish between \(k=0\) and \(k=1\), followed by a nearly linear increase in MAE for \(k=[1\dots7]\).
For \(k > 7\) it can be seen that the classification networks have learned the maximum of \(k\) across the dataset, hence the prediction error decreases when \(k\) reaches its maximum.
This is because classification based models intrinsically have access to the maximum number of sources determined by the output vector dimensionality.

Figure~\ref{fig:ssec:exp_fixed_gains/A} indicates that \emph{Classification} outperforms the other two distributions while Poisson regression performs better than Gaussian regression which confirms the findings made in~\cite{Rezatofigh16} on object counts.
With respect to the input representation, as shown in  Figure~\ref{fig:ssec:exp_fixed_gains/B}, despite its larger input dimension, choosing linear STFT as generally results in a better performance compared to \emph{MEL40}, \emph{LOGSTFT} or \emph{MFCC20}.
\par
A detailed analysis of all distribution and feature combinations, not shown here due to space constraints, reveals that STFT + \emph{Classification} performs best. This model achieves results of (MAE \(0.38 \pm 0.28\)) for \(k=[0\dots10]\) while the \emph{VQ} baseline (MAE \(2.41 \pm{1.08}\)) only performs slightly better than a mean estimator predicting \(\hat{k} = 5\) (MAE \(2.73 \pm{1.63}\)).
To show the level of overestimation or underestimation, we depict all responses in a confusion matrix (see Figure~\ref{fig:confusion}). Unlike humans that generally tend to underestimate for the task of speaker count estimation~\cite{kawashima15}, one can see that our proposed model slightly overestimates for smaller \(k\).
\section{Conclusion and Outlook}%
\label{sec:conclusion}
We introduced the task of estimating the maximum number of concurrent speakers in a simulated ``cocktail-party'' environment using a data-driven approach.
We evaluated three different methods to output integer source count estimates in conjunction with defining cost function over which to optimize.
Our experiments revealed a tradeoff between better overall performance but requiring the maximum number of speakers to be estimated as prior knowledge (classification) and slightly worse performance when treated as a regression problem using Poisson distribution.
Furthermore, we investigated and evaluated suitable input representations.
Our final proposed BLSTM based classification model achieves mean absolute error of less than 0.4 speakers for zero to ten speakers.
We think this first study on data-driven speaker count estimation opens the field to interesting and new research. Future work would be to evaluate and optimize other network structures such as convolutional neural networks and investigate the strategy a machine learning source count model pursues to solve the problem.
 \subsubsection*{Acknowledgments}
  The authors gratefully acknowledge the compute resources and support provided by the Erlangen Regional Computing Center (RRZE).
\newpage
\small
\bibliographystyle{abbrv}
\bibliography{references}

\end{document}